\documentclass[12pt,preprint]{aastex}

\def\msol{\hbox{\kern 0.20em $M_\odot$}}
\def\lsol{\hbox{\kern 0.20em $L_\odot$}}
\def\rsol{\hbox{\kern 0.20em $R_\odot$}}
\def\sr{\hbox{\kern 0.20em sr}}
\def\srmu{\hbox{\kern 0.20em sr$^{-1}$}}
 
\def\g{\hbox{\kern 0.20em g}}
\def\gmu{\hbox{\kern 0.20em g$^{-1}$}}
\def\kg{\hbox{\kern 0.20em kg}}
\def\pc{\hbox{\kern 0.20em pc}}
 
\def\mum{\hbox{\kern 0.20em $\mu$m}}
\def\mumd{\hbox{\kern 0.20em $\mu$m$^{-2}$}}
\def\cm{\hbox{\kern 0.20em cm}}
\def\m{\hbox{\kern 0.20em m}}
\def\km{\hbox{\kern 0.20em km}}
\def\nm{\hbox{\kern 0.20em nm}}
 
\def\s{\hbox{\kern 0.20em s}}
\def\h{\hbox{\kern 0.20em h}}
\def\sec{\hbox{\kern 0.20em sec}}
\def\min{\hbox {\kern 0.20em min}}
\def\smu{\hbox{\kern 0.20em s$^{-1}$}}
\def\smd{\hbox{\kern 0.20em s$^{-2}$}}
\def\an{\hbox{\kern 0.20em an}}
\def\anmu{\hbox{\kern 0.20em an$^{-1}$}}
\def\deg{\hbox{\kern 0.20em $^{\rm o}$}}
\def\yr{\hbox{\kern 0.20em yr}}
\def\yrmu{\hbox{\kern 0.20em yr$^{-1}$}}
\def\Myr{\hbox{\kern 0.20em Myr}}
\def\Mymu{\hbox{\kern 0.20em Myr$^{-1}$}}
 \def\K{\hbox{\kern 0.20em K}}
 \def\pcmu{\hbox{\kern 0.20em pc$^{-1}$}}
\def\pcmd{\hbox{\kern 0.20em pc$^{-2}$}}
\def\pcmt{\hbox{\kern 0.20em pc$^{-3}$}}
\def\kms{\hbox{\kern 0.20em km\kern 0.20em s$^{-1}$}}
\def\kmpd{\hbox{\kern 0.20em km$^{2}$}}
\def\kpc{\hbox{\kern 0.20em kpc}}
\def\cms{\hbox{\kern 0.20em cm\kern 0.20em s$^{-1}$}}
\def\erg{\hbox{\kern 0.20em erg}}
\def\ergs{\hbox{\kern 0.20em erg}}
\def\cmpd{\hbox{\kern 0.20em cm$^2$}}
\def\cmmd{\hbox{\kern 0.20em cm$^{-2}$}}
\def\cmms{\hbox{\kern 0.20em cm$^{-6}$}}
\def\cmpt{\hbox{\kern 0.20em cm$^3$}}
\def\cmmt{\hbox{\kern 0.20em cm$^{-3}$}}
\def\mpd{\hbox{\kern 0.20em m$^2$}}
\def\mmd{\hbox{\kern 0.20em m$^{-2}$}}
\def\mpt{\hbox{\kern 0.20em m$^3$}}
\def\mmt{\hbox{\kern 0.20em m$^{-3}$}}
\def\mujy{\hbox{\kern 0.20em $\mu$Jy}}
\def\mjy{\hbox{\kern 0.20em mJy}}
\def\Mj{\hbox{\kern 0.20em MJy}}
\def\jy{\hbox{\kern 0.20em Jy}}
\def\ghz{\hbox{\kern 0.20em GHz}}
\def\srmd{\hbox{\kern 0.20em sr$^{-1}$}}
\def \kms{km~$\rm{s}^{-1}$}
\def \cc{$\rm{cm}^{-3}$}
\def \mum{$\mu$m}

\def\G{\hbox{\kern 0.20em G}}

\def\htwo{\hbox{H${}_2$}}
\def\h13cop{\hbox{H$^{13}$CO$^{+}$}}

\slugcomment{To appear Spitzer Special Issue, ApJS}

\shorttitle{Is Cepheus E}
\shortauthors{Noriega-Crespo et al.}

\begin{document}

\newcommand{\jfourteen}{\hbox{$J=14\rightarrow 13$}}
 
\title{Is the Cepheus E Outflow driven by a Class 0 Protostar$?$}

\author{Alberto Noriega-Crespo\altaffilmark{1},
Amaya Moro-Martin\altaffilmark{2}, 
Sean Carey\altaffilmark{1},  
Patrick W. Morris\altaffilmark{1},
Deborah L. Padgett\altaffilmark{1}, 
William B. Latter\altaffilmark{1}, 
James Muzerolle\altaffilmark{2}}

\altaffiltext{1}{SPITZER Science Center, California Institute of Technology, CA 91125  USA}
\altaffiltext{2}{Steward Observatory, University of Arizona, 933 N Cherry Ave, AZ  85721 USA}

\begin{abstract}

New early release observations of the Cepheus E outflow 
and its embedded source, obtained with the Spitzer Space Telescope, 
are presented. We show the driving source is detected in all 4 IRAC bands, 
which suggests that traditional Class 0 classification, although 
essentially correct, needs to accommodate the new high sensitivity infrared
arrays and their ability to detected deeply embedded sources.
The IRAC, MIPS 24 and 70\mum~new photometric points are consistent with
a spectral energy distribution dominated by a cold, dense
envelope surrounding the protostar.
The Cep E outflow, unlike its more famous cousin the HH 46/47 outflow,
displays a very similar morphology in the near and mid-infrared
wavelengths, and is detected at 24\mum.
The interface between the dense molecular gas (where Cep E lies)
and less dense interstellar medium, is well traced by the emission
at 8 and 24 \mum, and is one of the most exotic features of the new IRAC and MIPS 
images. IRS observations of the North lobe of the flow confirm that
most of the emission is due to the excitation of pure H$_2$ rotational
transitions arising from a relatively cold (T$_{ex}\sim$ 700 K) and dense 
(N$_{H}\sim 9.6\times 10^{20}$ cm$^{-2}$) molecular gas.
\end{abstract}


\section{Introduction}

Class 0 objects were defined a decade ago (Andr\'e, Ward-Thompson \& Barsony 1993)
as those protostellar objects undergoing a gravitational collapse. They
are characterized by strong submillimeter continuum emission and relatively weak
or `non existent' infrared emission shortward of $\sim 10$\mum~(Andr\'e, Ward-Thompson,
\& Barsony 2000). 
Observationally, all the known Class 0 objects show traces of active mass loss in 
the form of outflows (Andr\'e, Ward-Thompson \& Barsony 2000), and therefore, 
theoretical schemes to describe their evolution invoke a delicate balance between mass 
accretion (to form the star)  and mass loss, to drive the outflow and remove angular 
momentum to sustain the star formation process 
(Smith 1998; Myers et al. 1998; Smith 2000; Froebrich et al. 2003).

The Cepheus E (Cep E) outflow and its driving source are considered a very good example
for studying this interface between the protostar and its mass loss, given the youth of its outflow, 
with a dynamical age of $\sim 5000$ yr, and the strength of its millimeter continuum emission, 
1 Jy at 1.3mm (Chini et al. 2001; Lefloch, Eisl\"oeffel \& Lazareff 1996). 
At a distance of 730 pc, Cep E-mm is one of the brightest Class 0 protostars known, 
and likely to become an intermediate ($\sim$3\msol) mass star 
(Moro-Martin et al. 2001; Froebrich et al. 2003).

Cep E was first identified as an outflow based on millimeter CO observations (Sargent 1977;
Fukui 1989),  followed by near infrared and higher spatial resolution CO studies
(Hodapp 1994; Eisl\"offel et al. 1996; Ladd \& Hodapp 1997; Ladd \& Howe 1997b; 
Noriega-Crespo, Garnavich \& Molinari 1998).
The outflow itself is quite compact, $\sim$1.5\arcmin, and driven by the IRAS 23011+6126
source, also known as Cep E-mm. The outflow is deeply embedded and nearly invisible
at visible wavelengths, with the exception of its South lobe which is breaking through the
molecular cloud (Noriega-Crespo 1997; Devine, Reipurth \& Bally 1997; Ayala et al. 2000;
Noriega-Crespo \& Garnavich 2001) and is named Herbig-Haro (HH) object 377. 
The properties of the outflow have been thoroughly analyzed in two recent
studies (Moro-Martin et al. 2001; Smith, Froebrich \& Eisl\"offel 2003). 
A comparative study of Cep E-mm source, in the context of other well known Class 0/I sources, 
was recently carried by  Froebrich et al. (2003).
We will rely on these works for the interpretation of the Spitzer observations.

In this {\it Letter} we present new data on the Cep E source and outflow
obtained with the Spitzer Space Telescope using its three instruments aboard. 
A brief description of the observations is presented in $\S$ 2,
in $\S$ 3 is the analysis of the data, and in $\S$ 4 we summarize
our results.

\section{Observations}
 
The Spitzer Cep E data were obtained during the in-orbit checkout
and Science Verification period. Cep E was observed using MIPS
(Rieke et al. 2004) on 2003 Nov 29, IRAC  (Fazio et al. 2004) 
on Nov 26 and IRS (Houck et al. 2004) on Nov 29.
The MIPS observations were performed in photometry mode at 24\mum~ and
fine scale 70\mum, with 3 and 10 sec frame times respectively, and
a total integration time of $\sim$48 and $167$ sec, accordingly.
The IRAC observations used high dynamic range 
12 sec frames with an on-source time of 108 sec. 
The IRS observations are of the North section of embedded outflow 
($\alpha$ = 23$^h$03$^m$13.6$^s$, $\delta$=+61\arcdeg42\arcmin43.5\arcsec~ J2000).
The spectroscopic observations were tuned to obtain a
high signal-to-noise ratio ($>$ 50) spectrum
over the 5.5 to 37\mum~wavelength range, based on our previous knowledge
of the Cep E mid-IR emission (Moro-Martin et al. 2001). Ramp 
durations of the different IRS modules were adjusted to achieve this.
The data from the 3 instruments were processed with the Spitzer Science
Center Pipeline S9.5\footnote{http://ssc.spitzer.caltech.edu/postbcd/}
which provides the basic calibrated data, 
the starting point for further analysis. The IRS spectra was analyzed using ISAP
\footnote{http://www.ipac.caltech.edu/iso/isap/isap.html}.

\section{Results and Discussion}

\subsection{The IRAC-MIPS Morphology and Photometry}

Figure 1 shows an image of Cep E in the near-IR vibrational line H$_2$ 
v = 1--0 S(1) at 2.12\mum~(Noriega-Crespo, Garnavich \& Molinari 1998) over a 2\arcmin~FOV;
the position of the IRAS 23011+6126 (Cep E-mm) is marked with a cross. A composite image 
using the four IRAC channels is shown in Figure 2. At first sight the morphology of outflow 
in the H$_2$ near infrared emission is remarkably similar to that at 3.6, 4.5, 5.7 and 8\mum.
Some noticeable differences are (i) the embedded  driving source is bright 
and undoubtly detected, (ii) there are
some fan structures at the base of both outflow lobes, (iii) the H$_2$ emission (green color) 
is the dominant component (see below), and (iv) the edge of the cloud , where HH 377 is visible, 
is traced by the 8\mum~emission (red color) and is probably related to 
the 7.7\mum~PAH feature.

A similar, and a bit more ambitious because of large wavelength range coverage, 
is the comparison shown in Fig. 3, where the H$_2$ 2.12\mum,
IRAC 4.5\mum~and MIPS 24\mum~emission are shown together. 
The comparison is relevant because it  shows the presence of the outflow at 24\mum, 
despite the very bright source, and a larger extension
of the warm dust at the edge of the cloud (red color).
We performed photometric measurements of the Cep E-mm source using 
IDP3\footnote{http://mips.as.arizona.edu/MIPS/IDP3/}, including PSF-fitting
with apertures $\sim 6\arcsec$ for the IRAC channels, and tuning the sky background measurements
to the extended emission from the outflow, which happens to be
insignificant in relationship with the flux of the source (see below).
The MIPS measurements at 24 and 70\mum~ include the outflow emission, but this contamination
is estimated to be less than 1\%. The integrated fluxes for the source are in Table 1, 
the formal uncertainties on the measurements are $\sim10\%$ for IRAC and MIPS 24\mum. For the 70\mum~
fine scale measurement the uncertainty is likely to be higher, 20\%, mostly because the source
is near the edge of the good part of array and we are underestimating the flux even after
the aperture correction.

The spectral energy distribution (SED) using previous and Spitzer measurements 
is shown in Fig. 4, and is compared also with a very simple SED model. 
The model has been described in detail elsewhere (Noriega-Crespo, Garnavich \& Molinari 1998).
It is based on an envelope made of spherically
symmetry dust shells where the temperature and density are assumed to follow a power law and
exponents of these mathematical relations are adjusted to fit the observations. 
The model in Fig. 4 has identical parameters as those in Moro-Martin et al. (2001), which 
uses silicates as the main component of the dust grains, and therefore, produces big absorption bands
at 10 and 20\mum. One of the reasons why we adopted this model was the fact that ISOCAM observations
did not detect the source at 9.6\mum~(Noriega-Crespo, Garnavich \& Molinari 1998).
The model SEDs in Fig. 4 are only meant to illustrate what is expected from an dust envelope
model and to bring into context the new IRAC and MIPS measurements.
There is the concern that the IRAC measurements
could be contaminated by shock emission from the outflow. In Cep E-mm source this certainly
could be the case if a large aperture ($>$ 6\arcsec) is used (see e.g. Engelke at al. 2004, Fig. 10).
Figure 5 shows the CVF ISOCAM spectrum of Cep E-mm, the spectrum was extracted from 
a single $6\arcsec~\times 6\arcsec$ pixel, and shows strong silicate, H$_2$O
and CO$_2$ absorption features, characteristic of a deeply embedded  source,
but not traces of H$_2$ emission from the outflow itself (Moro-Martin et al. 2001).
We believe, therefore, that any light contribution from the outflow in the Cep E-mm source 
IRAC measurements is negligible in this case.
The dust envelope SED model predicts a mass of $\sim$13\msol~ and 
a bolometric luminosity of $\sim$30\lsol.  SED models based on a graybody approximation 
(Froebrich et al. 2003) predict a factor two higher luminosity ($\sim$ 78\lsol) and temperature
($\sim$ 35 K), although the envelope mass differs only by 25\% ($\sim$8\msol) with respect
to our model.

We have pointed out before (Cernicharo et al. 2000) that despite the many magnitudes of
absorption produced by the dense envelopes surrounding Class 0 protostars and the
presence of deep absorption features from silicate dust and ices,
that is possible to detect low mass Class 0 protostars at shorter wavelengths 
than 10\mum, given the high sensitivity of infrared arrays in space. This is relevant
because of large surveys which are being performed by the Spitzer Legacy projects, aiming to
provide a complete census of the low end of the mass spectrum in nearby molecular clouds
(Evans et al. 2003), and should be taken in to account when a protostar class is assigned.

\subsection{\htwo\ Emission from Cep E North}

The IRS observations of the outflow were taken at the brightest region of the 
embedded redshifted lobe, as illustrated in Fig. 1. The
spectrum of Cep E North is dominated at short wavelengths ($<$13\mum) by 
the pure H$_2$ rotational lines (Fig. 6) as expected from previous studies
(Moro-Martin et al. 2001; Smith, Froebrich \& Eisl\"offel 2003; Noriega-Crespo 2002).
The detection of the H$_2$ 0-0 S(0) 28.22\mum~and 0-0 S(1) 17.04\mum~ lines 
is among the first ones in a low mass outflow, and demonstrates the superb 
sensitivity of the IRS spectrometer. Besides the H$_2$ rotational lines, two
atomic low excitation lines, [SI] at 25.25\mum~and [SiII] at 34.8\mum, are detected.
The spectrum shows the  11.3\mum~PAH band, and there are indications that
silicate absorption at 10\mum~and CO$_2$ ice at 15.2\mum~are also present.

The spectrum was dereddened using an E(B-V)$=1$, and the
measured H$_2$ integrated fluxes were used to estimate the mean excitation temperature
(assuming LTE) and the column density of the emitting molecular gas.
We found $T_{ex} = 686\pm 30$ K and N$_{H}\sim 9.6\times 10^{20}$ cm$^{-2}$, this column
density corresponds to a number density of $\sim 6.4\times 10^4$ \cc, assuming an
an envelope size of $\sim 1000$ AU (Moro-Martin et al. 2001). 

Studies of the mid and far-infrared emission from the Cep E outflow
(Moro-Martin et al. 2001; Smith Froebrich \& Eisl\'offel 2003) 
essentially agree with the scenario in which low velocity C-type shocks are 
the main excitation mechanism. Based on the relative strength of the H$_2$ lines 
and our estimate of the gas density, we derive a mean shock velocity of $\sim20$ \kms~
for the North outflow lobe (Kaufman \& Neufeld 1996). However,
the complex morphology of Cep E flow, which is made of multiple knots (Ladd \& Hodapp 1997),
may require as many as 20 C-type bowshocks to explain the integrated emission
of their lobes (Smith, Froebrich \& Eisl\"offel 2003).

\section{Summary}

We presented new observations of the Cep E outflow and its protostellar source,
using the 3 instruments aboard the Spitzer Space Telescope. We found that:

(1) The morphology of the outflow is remarkably similar to that of the near infrared
observations. Considering that Cep E provides the second example
of a low mass outflow observed with Spitzer (the other one being HH 46/47),
we can speculate that outflows from Class 0 sources do not develop a ``loop'' or 
a ``bubble'' structure (Noriega-Crespo et al. 2004), perhaps because they are
in the initial phase of breaking through  their dense placental envelope
and futher ejection events are required to develop a cavity.

(2) The Cep E-mm source or IRAS 23011+6126 was detected in all four IRAC channels.
Considering that the Cep E system lies at a distance of 730 pc, this suggests that
our traditional definition of a Class 0 source needs to accommodate the higher sensitivity
measurements, at wavelengths shorter than 10\mum, achievable with the new generation
of infrared arrays in space.

(3) The IRAC and MIPS integrated fluxes of Cep E-mm source are consistent with
the Class 0 envelope models, although in detail our simple SEDs models do not fit the
IRAC points. A comparison of the 5.7 and 8\mum~IRAC measurements with the Cep E-mm 
ISOCAM CVF spectrum shows a very good agreement.

(4) The IRS spectroscopic observations of the redshifted North lobe show a rich
spectrum of H$_2$ pure rotational lines, including strong  S(1) 17.04 and S(0) 28.22\mum~
lines. Assuming LTE we used these lines to estimate an
excitation temperature of $\sim700$K and a column density of 
N$_{H}\sim 9.6\times 10^{20}$ cm$^{-2}$ for a collisionally excited molecular gas.
A comparison with C-type shock models, using the H$_2$ line ratios and a gas
density of $10^5$ \cc, gives a mean shock velocity of 20\kms, consistent with
previous estimates.

\begin{acknowledgements}
This work is based on observations made with the Spitzer Space Telescope, 
which is operated by the Jet Propulsion Laboratory, California Institute
of Technology under NASA contract 1407. ANC's research was partially supported 
by NASA-APD Grant NRA0001-ADP-096. AAH is supported by the Spanish
Program Nacional de Astronom{\'{i}}a y Astrof{\'{i}}sica under grant 
AYA2002-01055.
\end{acknowledgements}

\clearpage

\begin{deluxetable}{lcccccc}
\tablecaption{Integrated mid and far-infrared fluxes$^a$ of the Cep E-mm
source for the 4 IRAC channels, MIPS 24 and (fine scale) 70\mum~bands.}
\tablewidth{0pt}
\tablehead{
\colhead{Wavelength$^b$} & \colhead{3.5} & \colhead{4.6} & \colhead{5.7} & \colhead{8} & \colhead{24} & \colhead{70}}
\startdata
 & 4.5  & 29 & 75  & 140 & 3600 & 30500  \\
\enddata
\tablenotetext{a}{in milliJansky}
\tablenotetext{b}{Bandpass reference wavelength in \mum}
\end{deluxetable} 

\clearpage

\begin{figure}
\includegraphics[width=220pt,height=204pt,angle=0.]{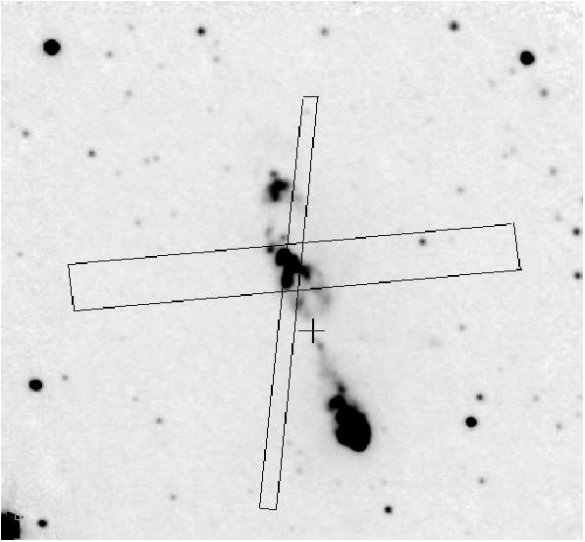}
\caption{\label{fig:1} Cep E outflow in H$_2$ at 2.12\mum~. The image shows
the schematic position of the Cep E-mm source (cross) and the Short-Low
and Long-low IRS slits. The map shows a 2\arcmin~$\times$2\arcmin~field. 
N is up and E is to the left.}
\end{figure}

\clearpage

\vspace{17cm}
\begin{figure}
\includegraphics[width=400pt,height=500pt,angle=0.]{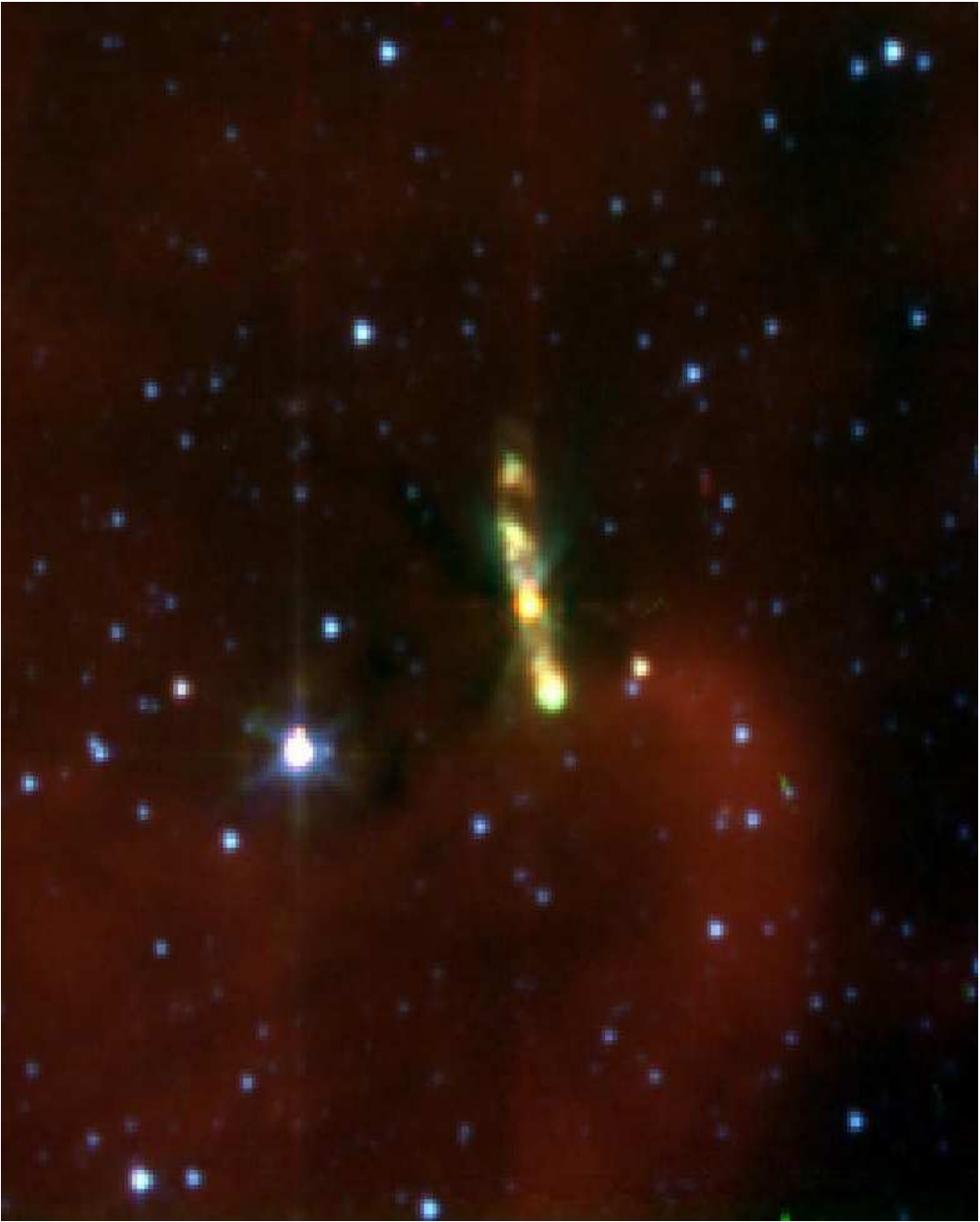}
\caption{\label{fig:2}[Color Plate] Spitzer image of the Cepheus E  embedded outflow
obtained using IRAC; the color scheme is the following:
3.6\mum~(blue), 4.5+5.8\mum~(green) and 8.0\mum~(red). 
The image covers a region of $\sim 5\arcmin  \times 5\arcmin$, and oriented
with North up and East to the left.}
\end{figure}

\clearpage

\vspace{17cm}
\begin{figure}
\includegraphics[width=400pt,height=500pt,angle=0.]{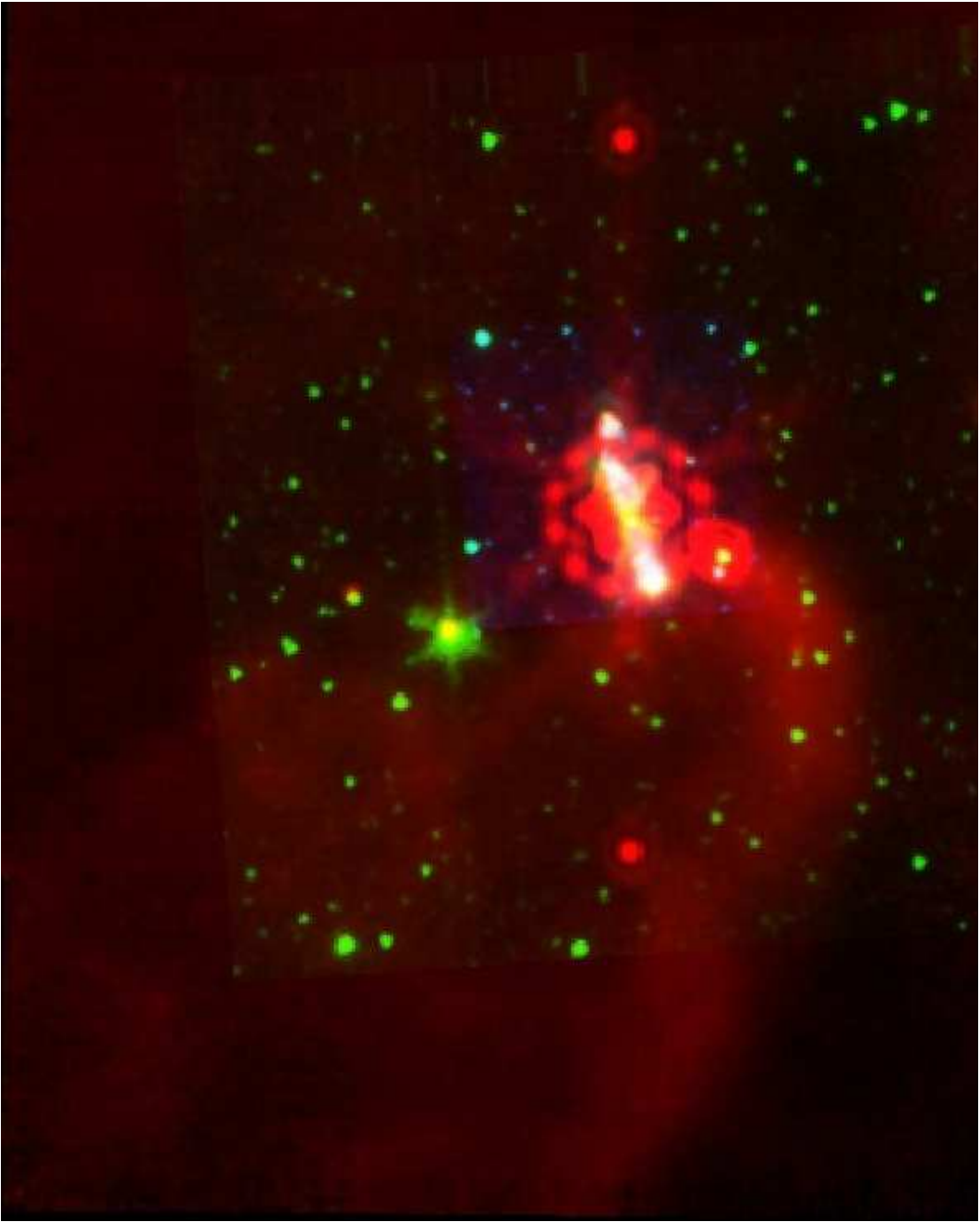}
\caption{\label{fig:3}[Color Plate] Spitzer image of the Cepheus E  embedded outflow
obtained using H$_2$ at 2.12\mum~(blue), IRAC Channel 2 at 4.5\mum~(green) and MIPS 24\mum~(red).
The image covers a region of $\sim 5\arcmin \times 7\arcmin$. North is up and East to the left}
\end{figure}

\clearpage

\begin{figure}
\includegraphics[width=200pt,height=240pt,angle=90.]{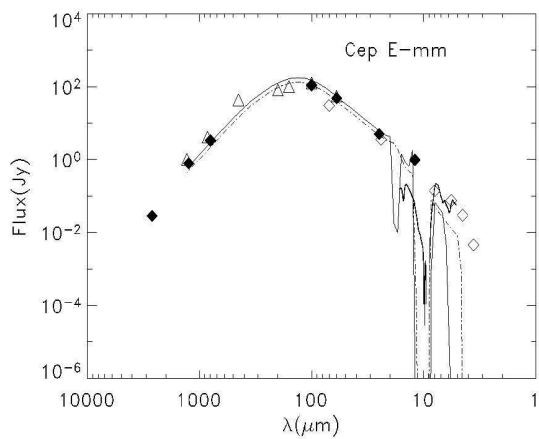}
\caption{\label{fig:4} Spectral energy distribution of the IRAS 23011+6126
source. Data points are taken  from Ladd \& Howe 1997 and Noriega-Crespo et al. 1998 
(filled diamonds), Froebrich et al. 2003 (open triangles) and
IRAC and MIPS measurements in this work (open diamonds). The CVF spectra of Cep E
is also included (thick solid line). Two simple models are used for comparison,
(1) a model assuming a dust opacity dominated by bare silicates, at a temperature of 18 K and 
$n = 7.5\times 10^4$ \cc~(solid line), and (2) a model with $n = 6.4\times 10^4$ \cc~
(broken line), 18 K and dust silicate with a thin ice mantle (Moro-Martin et. al 2001).}
\end{figure}

\begin{figure}
\includegraphics[width=200pt,height=230pt,angle=90.]{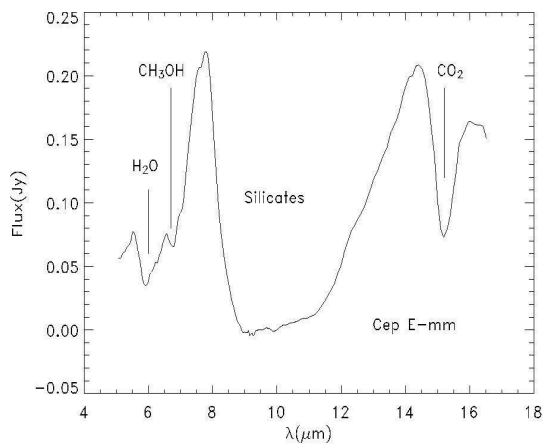}
\caption{\label{fig:5} Spectrum of Cep E-mm obtained with
ISOCAM Circular Variable Filter mode, with some strong
silicate and ice absorption  features (Moro-Martin et al. 2001).}
\end{figure}

\begin{figure}
\includegraphics[width=220pt,height=260pt,angle=90.]{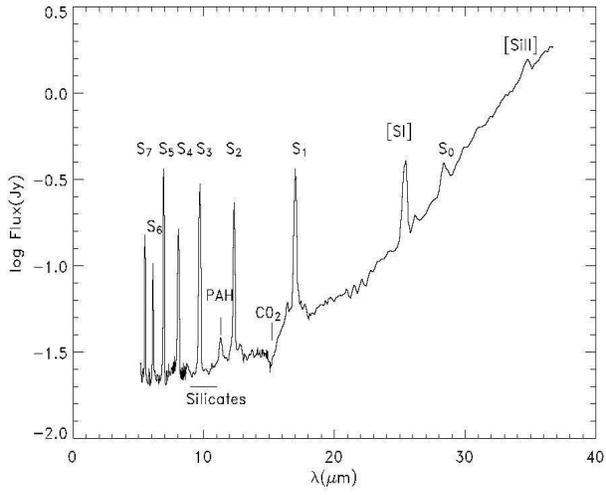}
\caption{\label{fig:6} IRS spectrum of the North lobe of the Cepheus E embedded outflow.}
\end{figure}

\end{document}